\documentclass[prl,twocolumn,amsmath,amssymb,aps]{revtex4}

\usepackage{graphicx}

\newcommand{\naught}{\mbox{\scriptsize o}}
\newcommand{\bigB}{\mathcal{B}}
\newcommand{\mword}[1]{\mbox{\scriptsize #1}}
\newcommand{\tmword}[1]{\mbox{\tiny #1}}

\newcommand{\Bext}{\bigB_{\mword{ext}}}

\newcommand{\BF}{\bigB_{\mword{F}}}
\newcommand{\Bflip}{\bigB_{\mword{flip}}}

\begin{document}

\title{Magnetoresistive Effects in Ferromagnet-Superconductor Multilayers}
\date{\today}
\author{E.M.\ Stoudenmire}
\author{C.A.R.\ S\'{a} de Melo}
\affiliation{Georgia Institute of Technology}

\begin{abstract}
We consider a nanoscale system consisting of Manganite-ferromagnet and Cuprate-superconductor 
multilayers in a spin valve configuration. The magnetization of
the bottom Manganite-ferromagnet is pinned by a Manganite-antiferromagnet. The
magnetization of the top Manganite-ferromagnet is coupled to the bottom one via indirect exchange
through the superconducting layers. We study the behavior of the critical temperature
and the magnetoresistance as a function of an externally applied parallel magnetic
field, when the number of Cuprate-superconductor layers are changed. There are two typical
behaviors in the case of a few monolayers of the Cuprates: a) For small magnetic fields, the
critical temperature and the magnetoresistance change abruptly when the flipping field
of the top Manganite-ferromagnet is reached. b) For large magnetic fields, the multilayered
system re-enters the zero-resistance (superconducting) state after having become
resistive (normal).
\end{abstract}

\maketitle

Magnetoresistive effects in ferromagnet/normal-metal multilayers have been
studied since the early 90's, where oscillations in exchange coupling and
magnetoresistance have been observed~\cite{parkin-90}. In this manuscript, we are interested in
magnetoresistive effects in the spin-valve configuration of ferromagnet/superconductor
multilayers. A spin-valve system consists of a pair of ferromagnetic layers anti-ferromagnetically
coupled and separated by a metal spacer where one layer is ``pinned,'' or fixed
in magnetization by an adjacent anti-ferromagnetic layer. Since electrons moving in the metal 
spacer encounter less resistance if the magnetizations of the layers are aligned, a spin-valve 
system can exhibit a magnetoresistance effect when an external magnetic field is applied
that is strong enough to flip the unpinned layer.
If ordinary low critical temperature ($T_{c}$) superconducting metals are used as spacers of 
thickness less than a $130$\AA, the proximity effect due to many lattice matched ferromagnets 
will be so strong as to destroy superconductivity, and thus any magnetoresistive effect associated with 
a superconducting metal spacer will be absent~\cite{sademelo-97}. If one insists on using 
a low $T_c$ superconducting metal it is necessary to increase its thickness more than $130$\AA\  
and use a weaker lattice matched ferromagnet. This situation was investigated
experimentally in CuNi/Nb multilayers, where a very weak magnetoresistive
effect was found~\cite{bader-02}. The ferromagnetic proximity effect is reasonably small such that it preserves
superconductivity, but the magnetoresistive effect is also small as the Nb spacer involves many
monolayers~\cite{bader-02}.  However, if the spacer layers are of thickness less than $130$\AA\ and consist
of metal Copper oxides like the high-$T_c$ d-wave superconductors
and the ferromagnets consist of Manganese oxides (CMR-materials), then
the ferromagnets can antiferromagnetically couple through the superconductor
given that the proximity effect due to ferromagnets is not
strong enough to suppress the superconducting state~\cite{sademelo-03}. Since it
has been experimentally demonstrated that a single monolayer of a Copper oxide can be superconducting
with a high $T_c$~\cite{bozovic-03}, and that magnetic coupling between CMR-ferromagnets separated by a complex
oxide is possible~\cite{goldman-00}, we propose in this manuscript that the critical temperature and the resistance
for a CMR-ferromagnet/high-$T_c$-superconductor multilayer
in such a spin-valve configuration can change dramatically as a function of
applied magnetic field.

Previous theoretical works that studied proximity and magnetoresistive effects
in F/S/F multilayers have focused on standard ferromagnets and standard superconductors and have used continuum
Usadel equations~\cite{tagirov-99,buzdin-99}.
Unfortunately, magnetoresistive effects in these systems have been demonstrated
experimentally to be small~\cite{bader-02}.
By contrast, we proceed by analyzing the dependence of the superconducting
critical temperature $T_{c}$ of such Manganese-oxide-ferromagnet-Copper-oxide-superconductor spin-valve systems
as a function of externally applied magnetic field.
First, we assume that the Curie temperature of the ferromagnets $T_{F} \gg T_{c}
$, where $T_{c}$ is the critical temperature of the superconductor spacer. Typical
values are $T_F \approx 300{\rm K}$ and $T_c \approx 85{\rm K}$. This condition implies that
the magnetization of each ferromagnet is essentially saturated when $T \approx
T_{c}$, and thus that the magnetic part of the free energy is essentially independent of
temperature for $T \approx T_{c}$.  We use a Ginzburg-Landau model (which can be derived rigorously for a d-wave
superconductor) to write the free energy density of each of the $n$ superconducting layers in terms of
the superconducting order parameter $\Psi(\mathbf{r})$.
The free energy for the $j^{\mword{th}}$ superconducting layer is
\begin{eqnarray}
\lefteqn{\mathcal{F}_{sj} = \mathcal{F}_{oj} } \nonumber \\
& & \mbox{} + \int dA \left[ \alpha_{j} \left|\Psi_{j}\right|^{2} +
\frac{\beta_{j}}{2} \left|\Psi_{j}\right|^{4}
+ \gamma_{j} \left|\nabla\Psi_{j}\right|^{2} \right] \label{energy_prelim},
\end{eqnarray}
%
where $\beta_{j} > 0$,  
$\alpha_{j}  =  \alpha_{\naught j} (T - T_{cj}^{\tmword{(o)}}) \label{alpha}$
and $T_{cj}^{\tmword{(o)}}$ is the
critical temperature of the superconducting material comprising an isolated
layer $j$.

\begin{figure}
\includegraphics{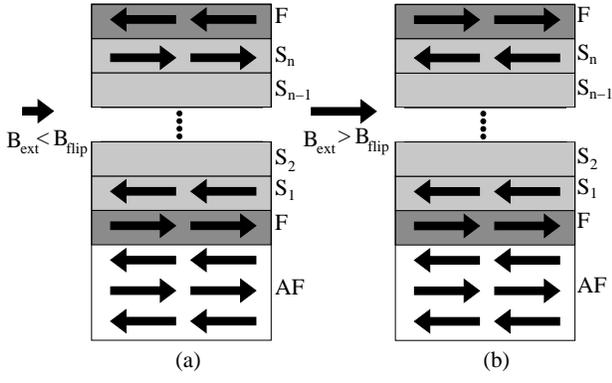}
\caption{\label{layers_schematic} Schematic of the ferromagnet magnetizations
and the magnetic fields felt by the superconducting
layers for
(a) $\Bext < \Bflip$ and
(b) $\Bext > \Bflip$.
}
\end{figure}

Since we will also be considering the effect of a uniform magnetic field
applied parallel to the plane of each layer, we add a term $\delta_{j} [\bigB_{\mword{tot}\:j}]^{2} |\Psi_{j}|^
{2}$ to each superconductor's free energy.  This term accounts for the coupling with the spin degrees of freedom of the
superconductor.  Note that $\delta_{j} > 0$ since a magnetic field introduces an
energy cost related to the superconducting electrons, which are paired in
singlet states.  Also note that the coupling of the parallel magnetic field with the charge
degrees of freedom is not relevant for small magnetic fields as the thickness of the superconducting spacer is
less than $130$\AA.  We assume that the ferromagnets surrounding the superconducting spacer are
identical, producing an exchange field of magnitude
$\BF$ felt only by neighboring superconducting layers (the ferromagnetic
proximity effect), and that the external field is applied
such that it is aligned with the magnetization of the pinned layer.
For the layer closest to pinned ferromagnet (see fig.~\ref{layers_schematic}),
then
$
\bigB_{\mword{tot}\:1} = \Bext - \BF \label{Btot1}
$
while all interior layers have simply $\bigB_{\mword{tot}\:j} = \Bext$
where $j = 2\ldots(n-1)$.
Most importantly, though, if $\Bflip$ is the critical applied field value
that flips the magnetization of the unpinned layer,
the field felt by the superconducting layer adjacent
to the unpinned ferromagnet is
\begin{equation}
\bigB_{\mword{tot}\:n} = \left\{ \begin{array}{ll}
                \Bext + \BF & \Bext < \Bflip \\
                \Bext - \BF  &  \Bext > \Bflip
                \end{array}
        \right. \label{Btotn}
\end{equation}
For the systems we will analyze, $\Bflip \ll \BF$ which is crucial in that a
small applied field may greatly affect the resistance of the system.
Since we are considering parallel magnetic fields (with no components
perpendicular to the superconducting layers),
we will take each superconducting order parameter $\Psi_{j}$ to be spatially
uniform, making the terms in (\ref{energy_prelim}) which depend on the gradient of $\Psi_
{j}$ zero.


To account for the Josephson coupling between superconducting layers,
we add terms of the form
$
\epsilon_{ij} \left|\Psi_{i} - \Psi_{j}\right|^{2}
$
to the free energy, but only consider coupling between adjacent layers.
Furthermore, we only consider the case where the
$\epsilon_{ij}$'s are nonnegative, i.e.\ there is no $\pi$-coupling as found when
superconducting layers are separated a ferromagnetic layer~\cite{chien-95}. 

To calculate the superconducting critical temperature of the system, taking all
superconducting layers to be identical,
we first write
$\tilde{\delta} = \delta/\alpha_{\naught}$ and $\tilde{\epsilon} = \epsilon/\alpha_{\naught}$,
and then seek the temperature for which $\mathcal{F}_{s}$ is no longer a local
minimum at $\mathbf{\Psi} = \mathbf{0}$.
For $\mathbf{\Psi}$ near zero, the free energy varies by an amount proportional
to
\begin{equation}
\sum_{ij} \alpha_{\naught}\,\Psi^{*}_{i}F_{ij}\Psi_{j} + \sum_{j} \beta_{j} \vert
\Psi_j \vert^4/2
\end{equation}
where the nonzero elements of the tridiagonal matrix $\hat{F}$ are given by
\begin{equation}
F_{ij} = \left\{ \begin{array}{ll}
                T - T^{\tmword{(o)}}_{c} + \tilde{\delta} [\bigB_{\mword{tot}
\,i}]^{2} \\
                \ \ \ + \tilde{\epsilon}_{i,\,i-1} + \tilde{\epsilon}_
{i,\,i+1} & \ i = j \\
                -\tilde{\epsilon}_{ij} & \ |i - j| = 1
                \end{array}
        \right.
\end{equation}
The system spontaneously enters the superconducting state only when $\hat{F}$
has negative eigenvalues, and so we may calculate the superconducting critical temperature of the system
by finding the largest value of $T$ that makes $\mbox{det}(\hat{F}) = 0$.


First consider only a single monolayer of a Copper oxide superconductor between
CMR-ferromagnets in the spin-valve configuration.
The critical temperature becomes, simply
%
%
\begin{equation}
T_{c} = T_{c}^{\tmword{(o)}} - \tilde{\delta}[\bigB_{\mword{tot}}]^{2} \label
{1layer_tc}
\end{equation}
\begin{figure}
\includegraphics{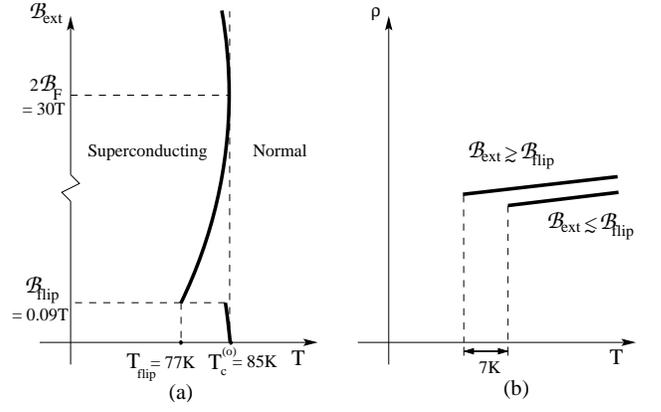}
\caption{\label{1layer_schematic} Schematic (a) phase diagram for a single YBCO
superconducting layer between LCMO ferromagnets and (b) the resistivity for current applied
parallel to the layers.  For the system depicted, $\tilde{\delta} \approx 1.6\,\mu_{B}^{2}/(T_{c}^
{\tmword{(o)}} k_{B}^{2})$.
Though the critical temperature drops by more than a few degrees as $\Bext$
increases beyond $\Bflip$, even larger drops may be attainable experimentally, depending on the choices of
the Copper oxide superconductor and CMR-ferromagnets.
}
\end{figure}
The magnetic proximity effect (exchange field present) in the superconducting
layer has its origins in the superexchange effect
between Manganese and Copper ions through common oxygen atoms shared in the
perovskite structure of the multilayer. The superexchange mechanism produces an antiferromagnetic
coupling between the Manganese and Copper spins, and thus the exchange field produced in the superconducting
layer by each ferromagnet is always anti-aligned with the magnetization of the respective ferromagnet
(see fig.~\ref{layers_schematic}).  Now, while the ferromagnetic layers surrounding the superconductor are still
anti-aligned in magnetization ($\bigB_{\mword{ext}} < \bigB_{\mword{flip}}$), the
superconductor is unaffected by the ferromagnets, whose exchange fields cancel each other.  However, for $\bigB_{\mword{ext}} > \bigB_{\mword{flip}}$, the superconductor
feels the combined magnetic field of the aligned ferromagnets. For the single layer case, then
\begin{equation}
\bigB_{\mword{tot}} = \left\{ \begin{array}{ll}
                \Bext & \Bext < \Bflip \\
                \Bext - 2\BF  &  \Bext > \Bflip,
                \end{array}
        \right. \label{1layer_Btot}
\end{equation}
such that the phase diagram for the system behaves as in fig.~\ref{1layer_schematic},
%
showing a dramatic change in the critical temperature at $\Bext = \Bflip$. 
This
abrupt change in $T_c$ provides for 
a super-colossal
magneto-resistive effect, where the magneto-resistance of the spin-valve system can change by several orders of
magnitude from zero ($\Bext < \Bflip$) to metallic values ($\Bext > \Bflip$) 
for systems cooled to temperatures within a reasonably large range (10\% of $T_c$ or larger)
(see fig.~\ref{1layer_schematic}). Though a system cooled to
a temperature within this {\it window of opportunity} becomes normal
for fields slightly above $\Bflip$, our model also predicts that the system re-enters 
the superconducting state for a large enough applied field ($\Bext$ anywhere from
$\Bflip$ to $2\BF$, depending on the temperature of the system).


For two superconducting monolayers deposited between anti-aligned ferromagnetic
layers, the critical temperature is
%
%
%
\begin{equation}
T_{c} = (T_{1} + T_{2})/2 + \sqrt{\left((T_{1} - T_
{2})/2 \right)^{2} +
\tilde{\epsilon}_{\tmword{12}}^{2}}.
\end{equation}
where $T_{1,2} \equiv T_{c}^{\tmword{(o)}} - \tilde{\delta}\left[\bigB_{\mword{tot}
\,1,2}\right]^{2} -
\tilde{\epsilon}_{\tmword{12}}$.
\begin{figure}
\includegraphics{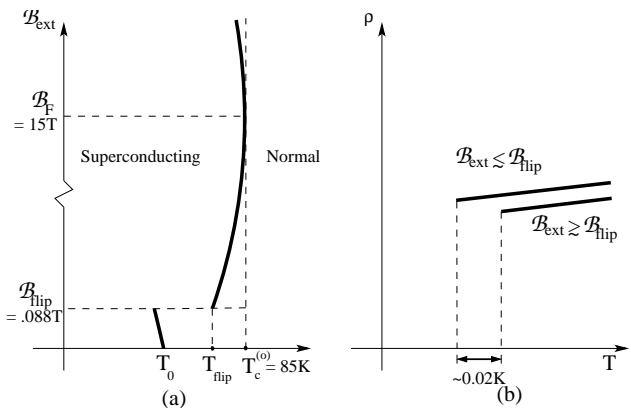}
\caption{\label{2layer_schematic} Schematic (a) phase diagram for two YBCO
superconducting layers between LCMO ferromagnets and (b) the resistivity for current applied
parallel to the layers.  For this particular system, $T_{c} = 82.99\,K$ for $\Bext$ slightly less that
$\Bflip$ and $T_{\mword{flip}} = 83.01\,K$.
The $\tilde{\delta}$ and $\BF$ parameters are the same as those used in
fig.~\ref{1layer_schematic}, but also $\tilde{\epsilon}_{\tmword{12}}$ has been
taken to be about 15\% of $T_{c}^{\tmword{(o)}}$, or $12.75\,K$.
}
\end{figure}

The phase diagram of a representative two-layer system is shown in fig.~\ref{2layer_schematic}.
While the two-layer system exhibits effects like those seen for the one-layer,
including the abrupt change in $T_{c}$ around $\Bflip$, there are some key differences
between the systems.

Even at zero field, the critical temperature of the two-layer system is less
than $T_{c}^{\tmword{(o)}}$ because a non-zero exchange field is already present in
each superconducting layer. Then, once $\Bext$ exceeds $\Bflip$, the change in $T_{c}$ is much
smaller than in the one-layer case because the magnitude of the field felt by the superconductor nearest to
the flipping ferromagnet changes only by $\Bflip$ in the two-layer case, while for the one-layer case the change
is very much larger ($2\BF$).  


Finally, two-layer systems whose temperature are in the window of opportunity
($T_{0} < T < T_{\mword{flip}}$) are normal first and then become
superconducting for $\Bext > \Bflip$, after which they remain superconducting
until the external field becomes extremely large ($> \BF$). Thus the superconducting-normal-
superconducting re-entrant behavior seen at relatively low fields for properly cooled one-layer systems is
basically no longer present in the two-layer system (except for
temperatures in a \emph{very} small range below $T_{0}$).

For systems having more than two superconducting monolayers, the behavior of $T_
{c}$ is qualitatively very similar that of the two-layer system. However, as the number of superconducting
layers is increased, the sign of magnetic coupling between ferromagnetic layers
starts to oscillate between ferromagnetic and antiferromagnetic~\cite{sademelo-03}. In addition, in the anti-aligned
cases, $\Bflip$ is reduced due to the weakened anti-ferromagnetic coupling between the ferromagnetic layers.
%
%
As the number of superconducting layers gets larger
there is essentially no difference in $T_c$ for parallel on anti-parallel orientations of the ferromagnets.
Furthermore, $T_c$ starts to reach its three-dimensional bulk value $T_c^{(3D)}$ as the effects of the ferromagnets
become confined to the {\it surfaces} of the superconductor, and thus become relatively speaking much weaker than in
the case of a few monolayers.

We would like to acknowledge support from NSF (DMR-0304380).
\bibliography{suplayer_bib}
\bibliographystyle{unsrt}

\end{document}